\documentclass[iop]{emulateapj}
\usepackage{amsmath}
\usepackage{amssymb}

\begin{document}

\title{Magnetohydrodynamic Fluid Stability in the Presence of \\Streaming Cosmic Rays}
\author{E. J. Greenfield\altaffilmark{1}, J. R. Jokipii\altaffilmark{2} and J. Giacalone\altaffilmark{2}}
\altaffiltext{1}{Department of Physics, University of Arizona, Tucson, AZ 85721}
\altaffiltext{2}{Department of Planetary Sciences, University of Arizona, Tucson, AZ 85721}

\begin{abstract}
We examine the effects of streaming cosmic rays upstream of a strong, parallel collisionless shock.  We include explicitly the inertia of the cosmic rays in our analysis, which was neglected in previous work.  For parameters relevant to the acceleration of cosmic rays at a supernova blast wave, we find {\it no} MHD fluid instability that would lead to the amplification of the magnetic field above that given by the compression at the shock.  We show how to recover, from our own analysis, the cosmic-ray-driven MHD fluid instability found by previous authors.  We conclude that including the inertia and dynamics of the cosmic rays keeps the system stable.  More over, the cosmic ray current leads to an additional Hall-like term in the magnetic evolution equation.  The implications of this paper for acceleration of galactic cosmic rays at supernova remnants are briefly discussed.
\end{abstract}

\maketitle

\section{Introduction}
Blast waves propagating out from supernova remnants (hereinafter SNR) are thought to be the source of most galactic cosmic rays (hereinafter GCR) up to an energy of $\sim10^{15}$ eV \citep{koy:95,allen:97,tan:98,aha:99,ber:03,vink:03}.  Because the observed GCR spectrum follows a power law with constant index up to this energy, diffusive shock acceleration (hereinafter DSA) is the most likely primary acceleration mechanism \citep{axf:77,bell:78,bland:78, jok:82, jok:87}.  For acceleration in a SNR, DSA at a quasi-parallel shock is only able to account for energies up to $\sim 10^{14}$ eV using the observed interstellar magnetic field and maximum scattering rate \citep{lag:83}.  In order to reach the knee at $10^{15}$ eV the acceleration occurring at such a shock must be more rapid.  In the case of DSA, this means the particles must be bound to the shock more effectively by the magnetic field.  In the case of a parallel shock, more rapid acceleration can be accomplished by increasing the magnitude of the magnetic field.  

\citet{bell:04} has suggested a magnetohydrodynamic (hereinafter MHD) fluid instability which amplifies a transverse perturbation upstream of the shock to $\delta B/B_0\sim 100$, where $B_0$ is the magnitude of the original field.  This cosmic ray current-driven instability (hereinafter CRCD) is triggered by GCRs streaming along the uniform background magnetic field.  This GCR current produces a separate return current in the background plasma which excites transverse turbulence in the upstream region.  

The sole purpose of the GCRs is to generate the return current which maintains charge neutrality.  The GCR inertia is not considered as we will explain below.  Once established, the excited transverse component scatters high energy GCRs, limiting their motion away from the shock increasing their acceleration rate.  He finds that the amplified field is adequate to accelerate GCRs up to, and possibly beyond, $10^{15}$ eV.  Observations of supernova remnants \citep{ber:03} strongly support the existence of an amplified magnetic field of the required magnitude.  Others have also worked in the kinetic regime to examine Bell's streaming instability.  While some work supports the large increase in $\delta B$ \citep{rev:06,blasi:08}, some find the amplification is only moderate $\delta B/B_0\lesssim10$ \citep{riq:09}.  Recently, \citet{riq:10} have proposed an additional instability caused by the {\it perpendicular} streaming of GCRs relative to the background field.  This perpendicular current-driven instability (hereinafter PDCI) is actually triggered by the pre-amplified field generated by the CRCD.  Additional work has also been carried out concerning kinetic instabilities related to self excited waves in shocks \citep{lee:83}.

An additional possibility to account for the observed acceleration to $10^{15}$eV is that much of the acceleration takes place where the shock is quasi-perpendicular, in which case no upstream magnetic-field amplification is necessary \citep{jok:82,jok:87}.  In that case, the observed amplification of the magnetic field could be the result of preexisting upstream turbulence warping the shock front, leading to turbulent motions of the plasma behind the shock that amplify the magnetic field through a turbulent dynamo process \citep{gia:07,ber:09}.  

In the present paper we will take a new approach to the original MHD problem of streaming GCRs at a quasi-parallel shock and compare the results with those found previously.  We find that if the inertia of the streaming cosmic rays is included and their dynamics considered, there is no MHD fluid instability.

\section{Linear Analysis}

\subsection{The CRCD MHD Fluid Instability}

It is well known that energetic particles accelerated at a quasi-parallel shock can stream ahead of the shock and excite waves in the upstream medium through kinetic processes.  Recently it has been proposed that these same particles can excite MHD fluid instabilities in the background fluid as well.  In this case, the shock-accelerated GCRs stream along the assumed uniform background magnetic field with a bulk speed $v_s$, the speed of the shock responsible for initially accelerating the GCRs.  As mentioned above this results in the amplification of magnetic perturbations transverse to the background magnetic field.

Beginning with the momentum equation for the background fluid, \cite{luc:00} and \cite{bell:04} find that the GCRs drive a MHD fluid instability in the background fluid.  Note that in this section and in direct references to \cite{bell:04}, results will be presented in SI units as in the original work.
\begin{align}
\rho \frac{d{\bf u}}{dt}&=-\nabla P + {\bf j}\times{\bf B}
\end{align}
Here $\rho$, $P$ and {\bf B} are the mass density, pressure and magnetic field of the background fluid.  The background fluid moves with velocity {\bf u} and the current in the background fluid is {\bf j}.  

The streaming GCRs are now introduced by way of Ampere's law\footnote{Note the use $\mu_0$ (due to use of SI units) in the above definition of the current.}, $\nabla\times{\bf B}=\mu_0({\bf j}+{\bf j}_{cr}-n_{cr}e{\bf u})$ where the RHS of this expression is the total current in the system.  Here ${\bf j}_{cr}$ is the current due to the streaming GCRs and the third term on the RHS preserves the quasi-neutrality of the system by providing an additional population of background electrons which cancel the positive charge of the GCRs.  The excess electrons are made to move with the same velocity as the background fluid, ${\bf u}$.  Substituting for the background current ${\bf j}$ using the above definition of the total current neglects the GCR inertia while including the forces acting on them and leads to the modified momentum equation for the background fluid.
\begin{align}
\rho\frac{d{\bf u}}{dt} =& -\nabla P\nonumber\\
&+\frac{1}{\mu_0}\left(\nabla\times{\bf B}\right)\times{\bf B}-{\bf j_{cr}}\times{\bf B}+ n_{cr}e{\bf u}\times{\bf B}\label{bellmom}
\end{align}
The second line of equation (\ref{bellmom}) is equal to the magnetic Lorentz force on the background plasma after solving for ${\bf j}$ in Ampere's law above.  The term $-{\bf j}_{cr}\times{\bf B}$ will ultimately drive the instability as discussed below.  At this point, we stress that the inertial term on the LHS of equation (\ref{bellmom}) represents only the inertia of the background fluid with mass density $\rho$ and velocity $\bf{u}$.  A corresponding inertial term for the GCRs (with variables $\rho_{cr}$ and ${\bf u}_{cr}$) is not present in this equation and only the electric current due to the GCRs has been retained.  This GCR current produces the return current in the background plasma which drives the instability.  In this way, the inertia of the GCRs has been neglected in Bell's treatment.  All subsequent discussions of inertia relate to this term.

In order to time-evolve the magnetic field in Bell's treatment, the assumption is made that the GCRs have no effect on the magnetic induction equation and the magnetic field is frozen into the {\it background} plasma as in ideal MHD.
\begin{align}
\frac{\partial{\bf B}}{\partial t}&=\nabla\times\left({\bf u}\times{\bf B}\right),\label{bellind}
\end{align}
Equations (\ref{bellmom}) and (\ref{bellind}) are then linearized using the expressions ${\bf B}={\bf B}_{||}+{\bf B}_{\bot}$, ${\bf u}={\bf u}_{\bot}$, ${\bf E}={\bf  E}_{\bot}$ and ${\bf j}_{cr}={\bf j}_{||}+{\bf j}_{\bot}$.  All values labeled $||$ are time-independent zeroth order terms oriented in the $z$ direction (taken to be the direction of the uniform magnetic field) and those labeled $\bot$ are first order perturbations in the $x-y$ plane which vary in time and space.  The gradient of the pressure is neglected at this point because it does not affect the linear study of the problem at hand.  Equations (\ref{bellmom}) and (\ref{bellind}) are then combined to eliminate ${\bf u}_{\bot}$ leaving a differential equation for the perturbed magnetic field and perturbed current (equation (3) in \cite{bell:04}).  
\begin{align}
\frac{\partial^2{\bf B}_{\bot}}{\partial t^2}&-v_a^2\frac{\partial^2{\bf B}_{\bot}}{\partial z^2}+\frac{j_{||}}{\rho v_s}{\bf B}_{||}\times\frac{\partial{\bf B}_{\bot}}{\partial t}\nonumber\\
&+\frac{B_{||}}{\rho}{\bf j}_{||}\times\frac{\partial{\bf B}_{\bot}}{\partial z}+\frac{B_{||}}{\rho}\frac{\partial{\bf j}_{\bot}}{\partial{z}}\times{\bf B}_{||}=0\label{bellalf}
\end{align}
The speeds $v_a$ and $v_s$ are the background Alfv\`{e}n and shock speeds, respectively.  The perturbed current is eliminated using the relationship ${\bf j}_{\bot}=(\sigma j_{||}/B_{||}){\bf B}_{\bot}$ where $\sigma$ is a unitless quantity defined below (equation (4) in \cite{bell:04}).  
\begin{align}
\sigma=&\frac{B_{||}}{j_{||}}e^2\int_0^\infty 2\pi p_{\bot}dp_{\bot}\int_{-\infty}^\infty dp_{||}\frac{v_{\bot}p_{\bot}}{\omega-\omega_g-kv_{||}}\label{sigma}\\ 
&\times\left[v_{||}\left(\frac{\partial f}{\partial p_{\bot}^2}-\frac{\partial f}{\partial p_{||}^2}\right)-\frac{\omega}{k}\frac{\partial f}{\partial p_{\bot}^2}\right]\nonumber
\end{align}
Here $e$, $p$, $v$, and $f$ are the GCR charge, GCR momentum, GCR velocity and the GCR distribution function, respectively.  This is the term used by \cite{bell:04} to couple the GCRs to the magnetic field though this effect will be neglected in the study of the CRCD instability (discussed in section \ref{bellinst}).  As before, $\bot$ and $||$ denote quantities perpendicular and parallel to the equilibrium magnetic field. 

Assuming an oscillatory solution for ${\bf B}_{\bot}$ in equation (\ref{bellalf}), $\sim exp(ikz-i\omega t)$, Bell finds the following dispersion relation (equation (7) in \cite{bell:04}):
\begin{align}
\omega^2-v_a^2k^2\pm\zeta v_s^2\frac{k}{r_{g1}}(1-\sigma_1)=0 \label{belldis}
\end{align}
Here, $\zeta$ is a unitless parameter quantifying the strength with which the GCRs drive the instability.  
\begin{align}
\zeta=\frac{j_{||}B_{||}r_{g1}}{\rho v_s^2}
\end{align}
This variable follows from the ${\bf j}_{cr}\times{\bf B}$ term in (\ref{bellmom}).  The subscript $1$ on $r_{g1}$, the GCR gyro radius, and $\sigma$ refer to those values calculated for GCRs with momentum $p_1$, the lower bound on the pre-existing CR spectrum.  The quantity $\sigma_1$ is related to but differs\footnote{See \cite{bell:04} Sections 3 and 4 for details} from $\sigma$ above.

The consequences of equation (\ref{belldis}) will be discussed in section \ref{bellinst}.

\subsection{Composite Fluid}\label{comp}

In our present treatment of this problem, {\it all} of the individual particle populations are treated as fluids which are then combined into a single fluid.  From this point forward, we refer to this single fluid as a composite fluid.  The inertia of the GCRs (here assumed to be singly charged with rest mass $m_p$) is explicitly included by considering the dynamical equation of the GCRs.  This is in contrast to the work of previous authors working in the MHD limit who have ignored the inertia of the GCRs and simply included the compensating return current in the momentum equation of the background fluid (as in equation (\ref{bellmom}) above).  We begin with four separate particle populations which are the background ions ($n_{bi}$), background electrons ($n_{be} = n_{bi}$), GCRs ($n_{cr}$) and the excess electrons required for quasi-neutrality ($n_{ex} = n_{cr}$).  The separation of the excess electrons from those in the background is not necessary but done here to emphasize the quasi-neutrality of the composite fluid.

\subsubsection{Composite Momentum Equation}

The composite momentum equation is found by adding together the momentum equations of the four populations.  The index $\alpha$ implies a sum over the four populations.
\begin{align}
&\sum_{\alpha=1}^4 \frac{d}{dt_{\alpha}}n_{\alpha}m_{\alpha}{\bf u}_{\alpha}=\sum_{\alpha=1}^4 n_{\alpha}e_{\alpha}\left(\frac{1}{c}{\bf u}_{\alpha}\times{\bf B}+{\bf E}\right)
\end{align}
For simplicity, we have taken the total plasma pressure to be isotropic.  Carrying out the summation gives the following composite momentum equation.
\begin{align}
n_{bi}m_p&\frac{d{\bf u}_{bi}}{dt}+n_{bi}m_e\frac{d{\bf u}_{be}}{dt}+\left<\gamma_{cr}\right>n_{cr}m_p\frac{d{\bf u}_{cr}}{dt}+n_{cr}m_e\frac{d{\bf u}_{ex}}{dt}\nonumber\\
&=\frac{e}{c}\left(n_{bi}{\bf u}_{bi}-n_{bi}{\bf u}_{be}+n_{cr}{\bf u}_{cr}-n_{cr}{\bf u}_{ex}\right)\times{\bf B}\label{firstmom}
\end{align}

The derivative used above is the total or convective derivative, $d/dt_{\alpha}\equiv\partial/\partial t+{\bf u_{\alpha}}\cdot\nabla$ where ${\bf u_{\alpha}}$ is the velocity of the fluid element being followed.  The velocity subscripts $bi$, $be$, $cr$, and $ex$ stand for background ions, background electrons, cosmics rays and excess electrons respectively.  The RHS of this equation is the total current, ${\bf j}$, crossed with the magnetic field.  The factor of $\left<\gamma_{cr}\right>$ in the above expressions arises due to the use of the relativistic momentum for the GCRs and represents an average Lorentz factor for the population of GCRs.  While the bulk speed of the GCRs, $u_{cr}$, is not relativistic, individual GCRs are relativistic in the shock frame while being accelerated by DSA.  The GCR fluid is essentially treated as a fluid of heavy ions enhancing their contribution to the momentum of the composite fluid.  Our work is not particularly sensitive to this quantity so we will simply treat $\left<\gamma_{cr}\right>$ as a constant and vary its value over a wide range of values.  A more in depth description of $\left<\gamma_{cr}\right>$ is outlined in Appendix \ref{appB}.  

Summing the left hand side as outlined in \cite{boyd:69}  (Appendix \ref{appA}) equation (\ref{firstmom}) simplifies to:
\begin{align}
\rho\frac{d{\bf u}}{dt}=\frac{1}{4\pi}(\nabla\times{\bf B})\times{\bf B}\label{mymom}
\end{align}
Here ${\bf u}$ is the center of mass velocity and $\rho$ is the total mass density of the composite fluid.  Note that the term on the LHS of equation (\ref{mymom}) includes the inertia of the GCRs.  Upon neglecting the contributions of the electrons to the fluid momentum ($m_e/m_i<<1$), ${\bf u}$ and ${\rho}$ can be expressed as below.
\begin{align}
{\bf u}=&\frac{n_{bi}{\bf u}_{bi}+\left<\gamma_{cr}\right>n_{cr}{\bf u}_{cr}}{n_{bi}+\left<\gamma_{cr}\right>n_{cr}}\label{totalu}
\end{align}
\begin{align}
\rho =& (n_{bi}+\left<\gamma_{cr}\right>n_{cr})m_p\nonumber\\
=&n_m m_p
\end{align}
We have substituted for the total current ${\bf j}$ using Ampere's law in the non-relativistic MHD approximation. 
\begin{align}
{\bf j} = \frac{c}{4\pi} \nabla\times{\bf B}
\end{align}

Including the dynamical equation of the GCRs from the beginning we avoid the more complicated momentum equation of Bell in equation (\ref{bellmom}) where the forces on the GCRs are included without their associated inertia.  Instead we find the usual momentum equation from ideal MHD.  

Due to the approximation of isotropic pressure the firehose instability will not appear in our analysis.  It should also be noted that the kinetic resonant streaming GCR instability \citep{kp:69} is also absent in our results due to our use of the MHD fluid equations.  \cite{bell:04} and \cite{rev:06} both note, however, that the growth of the non-resonant instability dominates that of the resonant instability.

\subsubsection{Composite Induction Equation}

Rather than assuming that the GCRs have no effect on the MHD induction equation, we explicitly determine the MHD electric field of this composite fluid \citep{boyd:03}.  To do so, we begin with the MHD momentum equation for the electrons.
\begin{align}
m_e&\frac{d}{dt}(n_{bi}{\bf u}_{be}+n_{cr}{\bf u}_{ex})\\
&=-\frac{e}{c}(n_{bi}{\bf u}_{be}+n_{cr}{\bf u}_{ex})\times{\bf B}-e(n_{bi}+n_{cr}){\bf E}\label{elecmom}\nonumber
\end{align}
Now, in the limit that $m_e<<m_i$ we neglect the the inertial term on the left hand side of equation (\ref{elecmom}).
\begin{align}
0\approx \frac{e}{c}(n_{bi}{\bf u}_{be}+n_{cr}{\bf u}_{ex})\times{\bf B}-e(n_{bi}+n_{cr}){\bf E}
\end{align}
Using the above equation we can solve for the electric field, {\bf E}, to find a generalized Ohm's Law.
\begin{align}
{\bf E}\approx&-\frac{1}{n_ec}(n_{bi}{\bf u}_{be}+n_{cr}{\bf u}_{ex})\times{\bf B}\label{myE1}
\end{align}
This shows the magnetic field is actually frozen into the total electron fluid as in the case of Hall MHD.  We define the variable $n_e\equiv n_{bi}+n_{cr}$.  We then eliminate the electron velocities using the definition of the {\it total} current, ${\bf j}=e(n_{bi}{\bf u}_{bi}+n_{cr}{\bf u}_{cr}-n_{bi}{\bf u}_{be}-n_{cr}{\bf u}_{ex})$.
\begin{align}
{\bf E}=-\frac{1}{c}\frac{n_{bi}}{n_e}{\bf u}_{bi}\times{\bf B}+\frac{1}{n_e ec}{\bf j}\times{\bf B}-\frac{1}{c}\frac{n_{cr}}{n_e}{\bf u}_{cr}\times{\bf B}\label{myE2}
\end{align}
By doing so we gain what we call an effective Hall term (the final two terms in equations (\ref{myE2})). The middle term is due to the transformation from the electron frame of reference to that of the ions and GCRs (center of mass) and is present in Hall MHD. The final term on the RHS of (\ref{myE2}) is due to the presence of the GCRs.    Physically, the Hall terms represents an electric field that is generated when particles of opposite charge separated due to the Lorentz force in a purely magnetic field; the Hall effect.  It can be thought of as a correction to the induction equation because the magnetic field is actually frozen into the electron fluid rather than the composite fluid moving at the center of mass velocity.  The correction accounts for the ions and GCRs having much larger gyro-radii than the electrons.  Lastly we express {\bf E} in terms of the center of mass velocity {\bf u}.
\begin{align}
{\bf E}=&-\frac{1}{n_e c}\left[n_m{\bf u} + (1-\left<\gamma_{cr}\right>)n_{cr}{\bf u}_{cr}\right]\times{\bf B}\nonumber\\
&+\frac{1}{4\pi n_e e}\left(\nabla\times{\bf B}\right)\times{\bf B}\label{myE3}
\end{align}
Inserting this expression for ${\bf E}$ into Faraday's law:
\begin{align}
\frac{\partial{\bf B}}{\partial t}=&-c\nabla\times{\bf E}\nonumber\\
\end{align}
 we arrive at the following induction equation for the evolution of the magnetic field.
\begin{align}
\frac{\partial{\bf B}}{\partial t}=&\nabla\times\left(\frac{n_m}{n_e}{\bf u}\times{\bf B}\right)\nonumber\\
&-\nabla\times\left[\frac{c}{4\pi n_{e}e}\left(\nabla\times{\bf B}\right)\times{\bf B}\right]\nonumber\\
&+\nabla\times\left(\frac{(1-\left<\gamma_{cr}\right>)n_{cr}}{n_e}{\bf u}_{cr}\times{\bf B}\right)\label{myind}
\end{align}
This differs from Bell's induction equation in (\ref{bellind}) where the magnetic field is assumed to be frozen into the background fluid.  There, the GCRs were assumed to have no effect on the form of the induction equation.  Equation (\ref{myind}) does not assume the magnetic field lines are frozen into the composite fluid and includes the effects of the GCRs.  The second and third terms on the RHS of equation (\ref{myind}) prevent the composite fluid from being well magnetized.  The only assumption necessary to arrive at this result is that $m_e/m_p << 1$.  Consideration of the relativistic mass of the GCRs leads to the second term in (\ref{myind}) because their contribution to the momentum of the fluid differs from their contribution to the current by a factor of $\left<\gamma_{cr}\right>$.  

\subsubsection{GCR Momentum Equation}

In order to find the perturbation to ${\bf u}_{cr}$ ($\delta u_{cr}$) in our subsequent linear analysis of equations (\ref{mymom}) and (\ref{myind}) we include the GCR momentum equation, (\ref{crmom2}) below.  This is necessary to provide a third equation for our system of three unknowns, $\delta B$, $\delta u$ and $\delta u_{cr}$.  This method differs from Bell's calculation of $\sigma_1$ from (\ref{sigma}) and was chosen as a more straight forward way to calculate the perturbations to the GCR velocity.  Because Bell takes $\sigma_1 \rightarrow 0$ to find his instability condition, the particular method for perturbing ${\bf u}_{cr}$ is not critical to our comparison with Bell.  What we consider to be important is retaining all effects due to the GCRs throughout our calculations.
\begin{align}
\left<\gamma_{cr}\right>n_{cr}m_p\frac{d{\bf u}_{cr}}{dt}=n_{cr}e\left(\frac{1}{c}{\bf u}_{cr}\times{\bf B}+{\bf E}\right)\label{crmom2}
\end{align}
The electric field can be eliminated by substituting ${\bf E}$ from equation (\ref{myE3}).
\begin{align}
\left<\gamma_{cr}\right>\frac{d{\bf u}_{cr}}{dt}=&\frac{n_m}{n_e}\frac{e}{m_p c}\left({\bf u}_{cr}-{\bf u}\right)\times{\bf B}\nonumber\\
&+\frac{1}{4\pi n_e m_p}\left(\nabla\times{\bf B}\right)\times{\bf B}\label{crmom}
\end{align}

\subsubsection{Perturbation Analysis}\label{pert}

We now perturb equations (\ref{mymom}), (\ref{myind}) and (\ref{crmom}) with transverse fluctuations in ${\bf B}$, ${\bf u}$ and ${\bf u}_{cr}$.
\begin{align}
{\bf B} &= \delta B(\hat{\bf x}\pm i\hat{\bf y})exp(ikz-i\omega t)+B_0\hat{\bf z}\label{bper}\\
{\bf u} &= \delta u(\hat{\bf x}\pm i\hat{\bf y})exp(ikz-i\omega t)+\left<\gamma_{cr}\right>\frac{n_{cr}}{n_m} u_{cr}\hat{\bf z}\label{uper}\\
{\bf u}_{cr} &= \delta u_{cr}(\hat{\bf x}\pm i\hat{\bf y})exp(ikz-i\omega t)+u_{cr}\hat{\bf z}\label{ucrper}
\end{align}
Equations (\ref{bper})-(\ref{ucrper}) assume the initial magnetic field to be uniform in the $\hat z$ direction and the streaming GCRs (also along $\hat z$) to be the only part of the composite fluid in motion.  Both $B_0$ and $u_{cr}$ are treated as constants.  Because of the coupling of perturbations in the $x$ and $y$ directions we have assumed the perturbations to be circularly-polarized.  The $\pm$ indicates the handedness of the perturbation with the $+$ being right-handed and the $-$ being left-handed.  The wave number $k$ is defined to be real while $\omega$ can be complex.  Both can take on positive or negative values.  

Inserting the assumed forms of {\bf B}, {\bf u} and ${\bf u}_{cr}$ from above into equations (\ref{mymom}), (\ref{myind}), and (\ref{crmom}) we find the following system of equations in our three unknowns.
\begin{align} 
&\left[\frac{n_{bi}}{n_m}\frac{kv_a^2}{B_0}\right]\delta B+\left[\omega-\left<\gamma_{cr}\right>\frac{n_{cr}}{n_m} k u_{cr}\right]\delta u=0\label{permom}
\end{align}
\begin{align}
&\left[\omega-\frac{n_{cr}}{n_e}k u_{cr}\mp\frac{n_{bi}}{n_e}\frac{k^2v_a^2}{\omega_g}\right]\delta B+\left[\frac{n_m}{n_e}kB_0\right]\delta u\nonumber\\
&+\left[(1-\left<\gamma_{cr}\right>)\frac{n_{cr}}{n_e}k B_0\right]\delta u_{cr}=0\label{perind}
\end{align}
\begin{align}
&\left[\frac{n_{bi}}{n_e}\frac{kv_a^2}{B_0}\mp \frac{n_{bi}}{n_e}\frac{\omega_g u_{cr}}{B_0}\right]\delta B\mp\left[\frac{n_m}{n_e}\omega_g\right]\delta u\nonumber\\
&+\left[\left<\gamma_{cr}\right>\omega-\left<\gamma_{cr}\right>ku_{cr}\pm\frac{n_m}{n_e}\omega_g\right]\delta u_{cr}=0\label{percr}
\end{align}
In the above equations (\ref{permom}) follow from (\ref{mymom}), (\ref{perind}) from (\ref{myind}) and (\ref{percr}) from (\ref{crmom}).  The variables $v_a$ and $\omega_g$ are the Alfv{\'e}n speed in the background plasma and gyro period of the background ions respectively.  Our assumption of transverse perturbations allows for the density parameters ($n_{m}$ and $n_{e}$) in equation (\ref{myind}) to remain constant in space and move outside the calculated curls.

This system of equations can be expressed as a matrix equation, $A\cdot x=0$.
\[\begin{pmatrix}
\alpha & \omega-\beta & 0\\
\omega-\gamma & \delta & \sigma\\
\Sigma & \kappa & \left<\gamma_{cr}\right>\omega-\chi\\
\end{pmatrix}
\begin{pmatrix}
\delta B\\
\delta u\\ 
\delta u_{cr}\end{pmatrix}=0\]
A solution for this system exists if the determinant of the coefficient matrix, $A$, vanishes.  The Greek variables in $A$ can be read from equations (\ref{permom})-(\ref{percr}).  Calculating this determinant gives a dispersion relation that is cubic in $\omega$.  While the roots of a cubic equation have analytic solutions, the complexity of the polynomial coefficients results in complicated roots better suited to numerical methods.  Rather than calculating the roots explicitly we will simply determine whether the roots will be purely real or complex over a wide range of the relevant parameters $k$, $\gamma_{cr}$, $u_{cr}$ and $n_{cr}$.  To do so we calculated the variables $Q$ and $R$.
\begin{align}
Q &\equiv \frac{a^2-3b}{9}\\
R &\equiv \frac{2a^3-9ab+27c}{54}
\end{align}
The variables $a$, $b$ and $c$ are the real coefficients of the monic cubic polynomial $\omega^3+a\omega^2+b\omega+c=0$.  If the quantity $Q^3-R^2<0$ the roots of the cubic will be complex and have a non-zero imaginary component \citep{press:92}.  In this case, the system would be unstable to perturbations.  This condition will be evaluated over the wide range of parameters listed below.
\begin{align}
B_0&=3\times10^{-6}\text{ G}\nonumber\\
n_{bi}&=1\text{ cm}^{-3}\nonumber\\
1&\leq\left<\gamma_{cr}\right>\leq10^6\nonumber\\
10^{-15}n_{bi}&<n_{cr}<10^{-1}n_{bi}\nonumber\\
20v_a&<u_{cr}<c/10\nonumber\\
-10^4k_{cr}&<k<10^4k_{cr}\nonumber
\end{align}  

The quantity $k_{cr}=2\pi/r_{g1}$ where $r_{g1} = 1.1\times 10^{18}$ cm from \cite{bell:04}.

In order to evaluate the condition for complex roots the values of $\left<\gamma_{cr}\right>$, $n_{cr}$, $u_{cr}$ and $k$ were varied over the ranges listed above by different step sizes.  Both $\left<\gamma_{cr}\right>$ and $n_{cr}$ were increased by powers of 10 and the value of $u_{cr}$ was increased by a factor of 20 with each successive calculation.  Lastly, the value of $k$ was incremented by $10^{-7}k_{cr}$.  

In order to check the accuracy of our calculation of the condition $Q^3-R^2<0$, $\left<\gamma_{cr}\right>$ was set equal to 1 while all other parameters were varied according to the process outlined above.  For this special case the roots of the dispersion relation are known to be real under all circumstances so $Q^3-R^2$ should never evaluate to a value $< 0$.  Upon running the calculation we find several cases that return $Q^3-R^2\sim -10^{-31}$, leading to false positives for complex roots.  Even if these complex roots were valid, the resulting growth rates are on the order of $10^{-12}$ s$^{-1}$ which are far too slow for the problem at hand.  It would take $10^4$ years to achieve $\delta{\bf B}/{\bf B}\sim 1$.  Using this information we adjust the condition for complex roots to be $Q^3-R^2< - 10^{-30}$ to account for numerical errors.  Under this corrected condition we find {\it all} roots of the dispersion relation to be real.  This means that there are {\it no} MHD fluid instabilities in our composite fluid formulation under reasonable conditions.  

\section{Comparison With Previous Work}

\subsection{The Original Cosmic-Ray-Driven MHD Instability}\label{bellinst}
Bell (2004) considers three separate regimes in which to examine the dispersion relation of equation (\ref{belldis}).  He states that two of the regimes are of little interest as the waves are only weakly driven by the GCRs.  These regimes are $kr_{g1}<1$ and $kr_{g1}>\zeta v_s^2/v_a^2$.  The intermediate regime, $1<kr_{g1}<\zeta v_s^2/v_a^2$, results in the GCR driven instability of interest here.  In this case, his dispersion relation simplifies to $\omega^2-v_a^2 k^2\pm\zeta v_s^2 k/r_{g1}$ (equation (15) of \cite{bell:04}) under the approximation that $\sigma_1\rightarrow 0$.  The choice of $+$ or $-$ is related to polarization and determines whether or not the mode will be oscillatory or growing.  A condition for the onset of the instability is readily found upon solving for $\omega$ in Bell's equation (15) above.  We have substituted for $\zeta$ in the equation below.
\begin{align}
\omega=v_ak\sqrt{1-\frac{\mu_0 j_{cr}}{kB_{||}}}
\end{align}
In order for the magnetic field to be unstable, the expression under the radical must be negative to give a complex $\omega$.  This requires a minimum value for $j_{cr}$.
\begin{align}
j_{cr}>\frac{kB_{||}}{\mu_0}\label{bellcon}
\end{align}
Working under the assumption that the condition in (\ref{bellcon}) is satisfied, Bell finds the maximum growth rate and corresponding wave number.  
\begin{align}
&\gamma_{max}=\frac{\zeta v_s^2}{2v_ar_{g1}}\label{kmax}
\end{align}
\begin{align}
&k_{max}=\frac{\zeta v_s^2}{2v_a^2r_{g1}}\label{bellgrow}
\end{align}
Recall that $\zeta$ represents the driving strength of the instability and follows from the ${\bf j}_{cr}\times{\bf B}$ term in (\ref{bellmom}) which only appears in (\ref{bellmom}) because the GCR current is introduced into the momentum equation of the background fluid without considering the GCR inertia.

\subsection{Simplifications of Composite Fluid to Recover Bell's Equations}
In this section we will outline the process to reduce our correct complete MHD equations to those found in \cite{bell:04}.  As we stated at the end of section \ref{comp}, the composite fluid formulation including the GCR inertia does not allow for MHD fluid instabilities to develop.  However, we can arrive at Bell's original system of equations if we do not properly account for the dynamics of the GCRs.  

We begin with the composite momentum equation (\ref{mymom}), which recall included the inertial terms for the background fluid, the excess electrons, and the GCRs.
\begin{align}
\rho\frac{d {\bf u}}{dt}=&\frac{1}{4\pi}\left(\nabla\times{\bf B}\right)\times{\bf B}\label{mymomagain}
\end{align}
In addition to the composite momentum equation, we reintroduce the momentum equation for the GCRs.
\begin{align}
\left<\gamma_{cr}\right>n_{cr}m_p\frac{d{\bf u}_{cr}}{dt} = n_{cr}e\left(\frac{1}{c}{\bf u}_{cr}\times{\bf B}+{\bf E}\right)\label{newcrmom}
\end{align}
Recall now equation (\ref{myE2}), the electric field found above subject to the condition of negligible electron inertia, $m_e<<m_i$.  To simplify our subsequent calculations we will now consider the limit that $n_{cr}/n_{bi} << 1$.  The middle term in (\ref{myE2}) is $\sim n_{cr}/n_{bi}$ and can now be dropped.  If we also neglect the Hall term (final term) from equation (\ref{myE2}) as done in \cite{bell:04}, the expression for the electric field simplifies to ${\bf E} = -(1/c){\bf u}_{bi}\times{\bf B}$.  Inserting this expression for ${\bf E}$ into equation (\ref{newcrmom}) the GCR momentum equation becomes the following.
\begin{align}
\left<\gamma_{cr}\right>n_{cr}m_p\frac{d{\bf u}_{cr}}{dt} = \frac{1}{c}{\bf j}_{cr}\times{\bf B}-\frac{1}{c}n_{cr}e{\bf u}_{bi}\times{\bf B}\label{newcrmom2}
\end{align}
Subtracting equation (\ref{newcrmom2}) from equation (\ref{mymomagain}) we find the equation below.
\begin{align}
\rho\frac{d {\bf u}}{dt}-&\left<\gamma_{cr}\right>n_{cr}m_p\frac{d{\bf u}_{cr}}{dt} \\
&= \frac{1}{4\pi}\left(\nabla\times{\bf B}\right)\times{\bf B}-\frac{1}{c}{\bf j}_{cr}\times{\bf B}+\frac{1}{c}n_{cr}e{\bf u}_{bi}\times{\bf B}\nonumber
\end{align}
The LHS of this equation reduces to the inertial term for the background plasma and extra electron population, with the electron inertia neglected as before.
\begin{align}
n_{bi}m_p\frac{d{\bf u}_{bi}}{dt} =& \frac{1}{4\pi}\left(\nabla\times{\bf B}\right)\times{\bf B}\nonumber\\
&-\frac{1}{c}{\bf j}_{cr}\times{\bf B}+\frac{1}{c}n_{cr}e{\bf u}_{bi}\times{\bf B}\label{newercrmom}
\end{align}
This equation is identical to Bell's momentum equation (equation (\ref{bellmom}) in this paper).  Note that ${\bf u}_{bi}$ in our formulation is the same as {\bf u} in Bell's work.  With ${\bf E} = -(1/c){\bf u}_{bi}\times{\bf B}$, the induction equation also reduces to that of \cite{bell:04}.
\begin{align}
\frac{\partial{\bf B}}{\partial t}&=\nabla\times\left({\bf u}_{bi}\times{\bf B}\right)\label{newind}
\end{align}
Equations (\ref{newercrmom}) and (\ref{newind}) represent two of the equations necessary for the development of the CRCD.  We have found these equations from our composite formulation by removing the dynamics of the GCRs.  Doing this has left out the GCR inertia from the above momentum equation but left information about the forces that act on them.  

In order to close the system of equations in this formalism, the dynamics of the GCRs must be accounted for in some way.  This was the purpose of $\sigma$ in \cite{bell:04}.  When $\sigma$ is set to $0$, however, the information about the GCR inertia is lost.  This would be equivalent to setting the LHS of equation (\ref{newcrmom2}) to zero, which implies that the final two terms on the RHS of equation (\ref{newercrmom}) are also zero, removing the mechanism for driving the CRCD.  

We suggest that if the inertia of the GCRs is retained in the study of this problem it should balance the Lorentz forces on the background fluid responsible for driving the CRCD.  In the event that the GCR inertia is neglected, the additional Lorentz forces are not consistent with the assumption that $\sigma \rightarrow 0$ and should also be set to $0$.  This conclusion is more apparent given our method for considering the GCRs instead of the quantity $\sigma$.

This section does not imply that we should find Bell's instability in the calculations of Section \ref{pert} for the case where $n_{cr}/n_{bi} << 1$.  The roots calculated in Section \ref{pert} were found for the case of the composite fluid where the GCR dynamics are included.  In this section, we retrieved Bell's equations by removing the GCR dynamics in addition to working in the limit of $n_{cr}/n_{bi} << 1$.

\subsection{Composite Formulation from Bell's Equations}
We can also retrieve our composite fluid momentum equation (\ref{mymom}) from Bell's momentum equation (\ref{bellmom}).  We begin with the momentum equation for the GCRs.
\begin{align}
\left<\gamma_{cr}\right>n_{cr}m_p\frac{d{\bf u}_{cr}}{dt} = n_{cr}e\left(\frac{1}{c}{\bf u}_{cr}\times{\bf B} + {\bf E}\right)\label{crmom3}
\end{align}
As we have stated before, the electric field in Bell's formulation is equal to: 
\begin{align}
{\bf E} = -\frac{1}{c}{\bf u}\times{\bf B}
\end{align}
Once again, ${\bf u}$ is the velocity of the background fluid.  Inserting this expression for ${\bf E}$ into (\ref{crmom3}) the momentum equation for the GCRs becomes the following.
\begin{align}
\left<\gamma_{cr}\right>n_{cr}m_p\frac{d{\bf u}_{cr}}{dt} = \frac{1}{c}{\bf j}_{cr}\times{\bf B} - \frac{1}{c}n_{cr}e{\bf u}\times{\bf B}
\end{align}
Adding this to equation (\ref{bellmom}) we find that all of the Lorentz force terms cancel leaving only the momentum equation of ideal MHD for the composite fluid.  This leaves no forces to drive an instability.  By accounting for the inertia and dynamics of the GCRs we find that no MHD instability is excited by the presence of the GCRs.

\subsection{Additional Work by Previous Authors}

As we mentioned, work on this topic has been carried out by a number of authors in both the MHD and kinetic regimes.  The purpose of this paper is to specifically address the previous linear MHD studies of the CRCD.  In the MHD regime, \cite{zir:08} have studied the behavior of the instability.  The authors confirm the existence of the instability proposed by Bell.  This is not unexpected, however, because the MHD equations used to describe the problem are identical to those of \cite{bell:04}.  In this case, the electric current resulting from the drift of the GCRs exerts a magnetic Lorentz force on the background plasma which drives the instability.  In this paper, we propose a different method for analyzing this problem, within the confines of MHD.  We have shown that if the inertia of the GCRs are treated on equal footing with that of the background plasma the relevant MHD equations take on a very different form.  Most noticeable is the lack of a magnetic Lorentz force in the momentum equation which would drive the instability.  We have shown that the total (magnetic and electric) Lorentz force responsible for driving the instability is exactly balanced by the MHD inertial term of the GCRs.  In the case of negligible GCR inertia, the total Lorentz force sums to zero leaving no net force to excite the instability.  

Kinetic and PIC studies have also been conducted with regards to the CRCD.  These simulations, with and without feedback on the GCRs, show the possibility for magnetic field amplification.  A number of such simulations do not consider parameters that include those stated in \cite{bell:04} or those that represent realistic SNR conditions.  \cite{gar:10} state that ``The diverse parameters chosen for the four sets of runs (A through D) do not directly reflect typical physical conditions found in young SNR shock precursors for all the runs."  Run A mentioned here most closely resembles the conditions of \cite{bell:04} but does not consider any feedback on the GCRs.  The study of \cite{niem:10} consider a highly relativistic {\it beam} of ions, not a slowly drifting and nearly isotropic distribution as we have done here.  The kinetic study by \cite{riq:09} also consider parameters very different from Bell.  In that case, the GCRs stream at velocities $> c/2$ where as Bell considered the case of GCRs streaming at a speed of c/30.  In addition, the ratio of $v_a/u_{cr}$ is $\sim 10^{-4}$ in \cite{bell:04} while \cite{riq:09} consider a ratio no less than $10^{-2}$.  Our range of parameters includes those originally used by \cite{bell:04} as well as those relevant to SNRs.  For this range we found no instability.  Direct comparison between our work and those numerical simulations above is difficult as the conditions in those papers consider GCR flow speeds approaching $c$.  The use of relativistic flow speeds is not valid in our non-relativistic MHD approximation.  This is why we have chosen to compare our results directly with those of the original MHD situation considered by Bell.

\section{Discussion and Conclusions}
In this paper we have considered the problem of MHD fluid stability in a plasma permeated by a fluid of streaming GCRs.  This work was carried out in the framework of MHD which is valid in typical astrophysical situations.  When considering the effects of the GCRs on the background plasma it is important to consider how the GCRs are treated in the problem.  Here we have discussed two different methods of approaching the problem which lead to very different results.

If the electric current created by the streaming GCRs is simply added to the dynamical equations for the background plasma without considering their own inertia, as it appears to have previously been done, there is a possibility for an instability to develop if condition (\ref{bellcon}) is satisfied.  The instability greatly enhances the previously zero transverse component of the magnetic field.  Physically, the instability is the result of the perturbed magnetic field being stretched out by the Lorentz force between the spiraling field and the neutralizing return current in the background plasma.  The $-{\bf j}_{cr}\times{\bf B}$ term in (\ref{bellmom}) acts as a driving force to what would otherwise be simple Alfv\`{e}n waves.  For a brief discussion of the physics of the instability as well as an illustrative figure of the relevant forces see Fig. 1 in \cite{zir:08}.  In the case of \cite{bell:04}, $j_{cr}$ is constant and free from any feedback allowing it to continually drive the return current and the instability.  

Here, we have chosen to re-examine the original MHD analysis while explicitly including the inertia and dynamics of the GCRs which has been neglected in previous MHD studies.  In this paper, we have shown that the method for including the GCRs has a direct bearing on the creation of an instability. As stated in \cite{niem:08}, ``Previously published MHD simulations have assumed a constant cosmic-ray current and no energy or momentum flux in the cosmic rays, which excludes a back-reaction of the generated magnetic field on cosmic rays, and thus the saturation of the field amplitude is artificially suppressed."  These authors found no evidence of the CRCD instability and only a moderate amplification of the magnetic field ($\delta B/B\sim 1$) due to turbulent motions in the plasma.  We have shown in this paper that failure to consider the dynamics of the GCRs leads to excess forces on the system (the ${\bf j}_{cr}\times{\bf B}$ term in (\ref{bellmom})) which stretch out the magnetic field lines.  This force is balanced by the GCR inertial in our composite fluid formulation.  

This paper highlights the importance of how the GCRs are accounted for in this problem.  By considering the dynamics of the GCRs (and thus their inertia) along with that of the background fluid we find that there is no such MHD fluid instability.  By working with the composite fluid, the MHD momentum equation contains the total current, {\bf j}, in the system rather than only the return current $-{\bf j_{cr}}$.  In the case of an equilibrium uniform magnetic field ${\bf j}=0$ so there is no driving term as in Bell's momentum equation (\ref{bellmom}) to excite an instability.  Physically, the lack of any net current means that there is no net Lorentz force to stretch out the perturbed magnetic field.  We have also shown that the CRCD is the result of neglecting the GCR inertia in the analysis of this problem.

The lack of an instability in our analysis calls into question our neglect of the kinetic resonant streaming instability due to its slow growth rate compared to that of the non-resonant instability.  This concern will be addressed through a numerical study of this problem in a subsequent paper.

The conclusion of this paper implies that the observed large magnetic field \citep{ber:03} in SNRs should be found in some process other than cosmic-ray-driven amplification upstream of the blast wave.  One such process is the downstream amplification caused by pre-existing upstream turbulence \citep{gia:07,ber:09}.  In this scenario, the acceleration to energies $\sim 3\times 10^{15}$eV could occur at the quasi-perpendicular parts of the blast wave.\footnote{Recall once again that in the case of DSA the original parallel shock would actually more closely resemble a perpendicular shock if $\delta B/B\sim100$.  As mentioned in the introduction, a perpendicular shock is capable of accelerating GCRs to higher energies without the need for this instability.}

\begin{acknowledgements}
This work was supported, in part, by NASA under grants NNX08AH55G and NNX10AF24G.  JRJ is grateful to A. R. Bell and M. A. Lee for helpful discussions which improved this paper.
\end{acknowledgements}

\appendix
\section{A. Derivation of Composite Fluid Momentum Equation}\label{appA}
The $ith$ component of the four momentum equations are shown below in index notation.  In order the equations are for the background ions, background electrons, GCRs and the neutralizing electrons.
\begin{align}
\frac{\partial}{\partial t}m_p n_{bi} u_{bi,i}+\frac{\partial}{\partial x_j}\left[m_p n_{bi}\left(u_{bi,i} u_j+u_{bi,j}u_i-u_i u_j\right)\right]=n_{bi}e\left[E_i+\frac{1}{c}\epsilon_{ijk}u_{bi,j}B_k\right]\label{ions}
\end{align}
\begin{align}
\frac{\partial}{\partial t}m_p n_{bi} u_{be,i}+\frac{\partial}{\partial x_j}\left[m_e n_{bi}\left(u_{be,i} u_j+u_{be,j}u_i-u_i u_j\right)\right]=-n_{bi}e\left[E_i+\frac{1}{c}\epsilon_{ijk}u_{be,j}B_k\right]\label{elec}
\end{align}
\begin{align}
\frac{\partial}{\partial t}\left<\gamma_{cr}\right>m_p n_{cr} u_{cr,i}+\frac{\partial}{\partial x_j}\left[\left<\gamma_{cr}\right>m_p n_{cr}\left(u_{cr,i} u_j+u_{cr,j}u_i-u_i u_j\right)\right]=n_{cr}e\left[E_i+\frac{1}{c}\epsilon_{ijk}u_{cr,j}B_k\right]\label{crs}
\end{align}
\begin{align}
\frac{\partial}{\partial t}m_p n_{cr} u_{ex,i}+\frac{\partial}{\partial x_j}\left[m_p n_{cr}\left(u_{ex,i} u_j+u_{ex,j}u_i-u_i u_j\right)\right]=-n_{cr}e\left[E_i+\frac{1}{c}\epsilon_{ijk}u_{ex,j}B_k\right]\label{exel}
\end{align}
Summing equations (\ref{ions})-(\ref{exel}) we arrive at the following equation for the composite fluid.
\begin{align}
\frac{\partial}{\partial t}(\rho u_i) + \frac{\partial}{\partial x_j}(\rho u_i u_j) = \frac{1}{c}\epsilon_{ijk}j_j B_k
\end{align}
Recall the definitions of $\rho$ and ${\bf B}$ from section \ref{comp}.  We then expand the derivatives on the LHS of the above equation.
\begin{align}
\rho\frac{\partial u_i}{\partial t} + \rho u_j\frac{u_i}{\partial x_j} + u_i\frac{\partial\rho}{\partial t} + u_i\frac{\partial\rho u_j}{\partial x_j} = \frac{1}{c}\epsilon_{ijk}j_j B_k
\end{align}
Factoring the third and fourth terms we see that these terms cancel by mass conservation.
\begin{align}
\rho\frac{\partial u_i}{\partial t} + \rho u_j\frac{u_i}{\partial x_j} + u_i\left[\frac{\partial\rho}{\partial t} + \frac{\partial\rho u_j}{\partial x_j}\right] = \frac{1}{c}\epsilon_{ijk}j_j B_k
\end{align}
We now substitute for the total current {\bf j} by Ampere's law to arrive at our momentum equation for the composite fluid, equation (\ref{mymom}) above.
\begin{align}
\rho\frac{d{\bf u}}{dt} = \frac{1}{4\pi}(\nabla\times{\bf B})\times{\bf B}
\end{align}

\section{B. Calculation of $\left<\gamma_{cr}\right>$}\label{appB}
The factor $\gamma_{cr}$ is an average Lorentz factor for the entire GCR population and can be calculated as shown below.  
\begin{align}
\left<\gamma_{cr}\right> = \frac{\int_{p_1}^{p_2}\gamma_{cr} f(p)dp}{\int_{p_1}^{p_2}f(p)dp}
\end{align}
If we assume the GCR distribution function $f(p)$ to be a power law, $\sim p^{-\Gamma}$, we find the following form for $\left<\gamma_{cr}\right>$.
\begin{align}
\left<\gamma_{cr}\right> \sim \frac{\int_{p_1}^{p_2}p^{-\Gamma}\left[\left(\frac{p}{m_p c}\right)^2+1\right]^{1/2}dp}{\int_{p_1}^{p_2}p^{-\Gamma}dp}
\end{align}
The limits of integration $p_1$ and $p_2$ are the minimum and maximum momentum of the CRs in the distribution function $f(p)$.  For $\Gamma=4$ as in the case of ideal DSA at a strong shock this integral can be solved analytically.
\begin{align}
\left<\gamma_{cr}\right>\sim \left[\left(\frac{p}{m_p c}\right)^2+1\right]^{3/2}\bigg|^{p_2}_{p_1}
\end{align}


\begin{thebibliography}{99}
\bibitem[Aharonian(1999)]{aha:99}
	Aharonian, F. A. 1999, Astropart. Phys., 11, 225
\bibitem[Aharonian et al.(2001)]{aha:01}
	Aharonian, F. A. et al. 2001, \aap, 370, 112
\bibitem[Allen et al.(1997)]{allen:97}
	Allen, G. et al. 1997, \apj, 487, L97
\bibitem[Amato and Blasi(2009)]{ama:2009}
	Amato, E. \& Blasi, P. 2009, \mnras, 392, 1591
\bibitem[Axford et al.(1977)]{axf:77}
	Axford, W. I. et al. 1977, Proc. 15th Int. Cosmic Ray Conf., 11, 132
\bibitem[Bell(1978)]{bell:78}
	Bell, A. R. 1978, \mnras, 182, 147
\bibitem[Bell(2004)]{bell:04}
	Bell, A. R. 2004, \mnras, 353, 550
\bibitem[Beresnyak et al.(2009)]{ber:09}
	Beresnyak, A. et al. 2009, \apj, 707, 1541
\bibitem[Berezhko \& Volk(2003)]{ber:03}
	Berezhko, E. G. et al. 2003, \aap, 412, L11
\bibitem[Blandford and Ostriker(1978)]{bland:78}
	Blandford, R. D. \& Ostriker, J. P. 1978, \apj, 221, L29
\bibitem[Blasi and Amato(2008)]{blasi:08}
	Blasi, P. \& Amato, E. 2008, Proc. 30th Int. Cosmic Ray Conf., 2, 235
\bibitem[Boyd and Sanderson(1969)]{boyd:69}
	Boyd, T. J. \& Sanderson, J. J. 1969, Plasma Dynamics
\bibitem[Boyd and Sanderson(2003)]{boyd:03}
	Boyd, T. J. \& Sanderson, J. J. 2003, The Physics of Plasmas, Cambridge University Press, Cambridge UK
\bibitem[Gargat\'e et al.(2010)]{gar:10}
	Gargat\'e, L. et al. 2010 \apj, 711, L127
\bibitem[Giacalone and Jokipii(2007)]{gia:07}
	Giacalone, J. \& Jokipii, J. R. 2007, \apj, 663, L41
\bibitem[Jokipii(1982)]{jok:82}
	Jokipii, J. R. 1982, \apj, 255, 716
\bibitem[Jokipii(1987)]{jok:87}
	Jokipii, J. R. 1987, \apj, 313, 842
\bibitem[Lucek and Bell(2000)]{luc:00}
	Lucek, S. G. \& Bell, A. R. 2000, \mnras, 314, 65
\bibitem[Koyama et al.(1995)]{koy:95}
	Koyama, K. et al. 1995, \nat, 378, 255
\bibitem[Krall and Trivelpiece(1973)]{krall:73}
	Krall, N. A. \& Trivelpiece, A. W. 1973, Principles of Plasma Physics. McGraw Hill, New York
\bibitem[Krymskii(1977)]{krym:77}
	Krymskii, G. F. 1977, Sov. Phys. Dokl., 23, 327
\bibitem[Kulsrud and Pearce(1969)]{kp:69}
	Kulsrud, R. \& Pearce, W. P. 1969, \apj, 156, 445
\bibitem[Lagage \& Cesarsky(1983)]{lag:83}
	Lagage, P. O. \& Cesarsky, C. J. 1983, \aap, 125, 249
\bibitem[Lee(1983)]{lee:83}
	Lee, M. A. 1983, \jgr, 87, 5063
\bibitem[Naito et al.(1999)]{naito:99}
	Naito, T. et al. 1999, Astron. Nachr., 320, 205
\bibitem[Niemiec et al.(2008)]{niem:08}
	Niemiec, J. et al. 2008, \apj, 684, 1174
\bibitem[Niemiec et al.(2010)]{niem:10}
	Niemiec, J. et al. 2010, \apj, 709, 1148
\bibitem[Press et al.(1992)]{press:92}
	Press, W. H. et al. 1992 Numerical Recipes in FORTRAN, 2nd Edition, Cambridge University Press, Cambridge UK
\bibitem[Reville et al.(2006)]{rev:06}
	Reville, M. A. et al. 2006, Plasma Phys. Control. Fusion, 48, 1741
\bibitem[Riquelme and Spitkovsky(2009)]{riq:09}
	Riquelme, M. A. \& Spitkovsky, A. 2009, \apj, 694, 626
\bibitem[Riquelme and Spitkovsky(2010)]{riq:10}
	Riquelme, M. A. \& Spitkovsky, A. 2010, \apj, 717, 1054
\bibitem[Stroman et al.(2009)]{stro:09}
	Stroman, T. 2009, \apj, 706, 28
\bibitem[Tanimori et al.(1998)]{tan:98}
	Tanimori, T. 1998, \apj, 497, L25
\bibitem[Vink and Laming(2003)]{vink:03}
	Vink, J. \& Laming, J. M. 2003, \apj, 584, 758
\bibitem[Zirakashvili et al.(2008)]{zir:08}
	Zirakashvili, V. N. et al. 2008, \apj, 678, 255
\end{thebibliography}
\end{document}